\documentclass[%
reprint,
 amsmath,amssymb,
 aps,
floatfix,
]{revtex4-2}

\usepackage{thmtools}
\usepackage{amsfonts}
\usepackage{amssymb}
\usepackage{graphicx}
\usepackage{dcolumn}
\usepackage{xcolor}
\usepackage{bm}
\usepackage{hyperref}
\usepackage{float}

\usepackage{subfigure} 

\begin{document}


\title{The inverse problem beyond two-body interaction: the cubic mean-field Ising model.}

\author{Pierluigi Contucci}
\email{pierluigi.contucci@unibo.it}
\affiliation{Dipartimento di Matematica, Università di Bologna, Italy}%

\author{Godwin Osabutey}
\email{godwin.osabutey2@unibo.it}
\affiliation{Dipartimento di Matematica, Università di Bologna, Italy}%
\author{Cecilia Vernia}
\email{cecilia.vernia@unimore.it}
\affiliation{%
 Dipartimento di Scienze Fisiche Informatiche e Matematiche, Università di Modena e Reggio Emilia, Modena, Italy
 }%


\date{\today}

\begin{abstract}
In this paper we solve the inverse problem for the cubic
mean-field Ising model. Starting from configuration data generated
according to the distribution of the model we reconstruct the free
parameters of the system. We test the robustness of this inversion 
procedure both 
in the region of uniqueness of the solutions and
in the region where multiple thermodynamics phases are present.
\end{abstract}

\maketitle


\section{Introduction}
In this paper we study the inverse problem for a class of mean-field models in statistical mechanics with cubic interaction. The direct problem of statistical mechanics is to compute macroscopic variables (i.e. the average values of magnetizations and correlations) when the couplings and fields are known. In the inverse problem the reverse is done: the couplings and fields are computed using the (statistical) datum of the macroscopic quantities. This technique is known sometimes as Boltzmann machine learning, a special case of learning in statistical inference theory~\cite{Jaynes1957, MacKay2003} when the probability measure is the Boltzmann-Gibbs one. 

In recent years, studies in deep learning for artificial intelligence have been approached in terms of inverse problem in statistical mechanics \cite{Nguyen_Zecchina_Berg_2017, Baldassi2018,Beentjes_2020}. The techniques to study that case are of very different nature than those we treat in this work because the parameters to be identified are of very high dimension and the involved models concern the theory of disordered systems \cite{Mezard_Parisi_Virasoro_1987}. Although in this study we are only interested in computing three parameters, we believe that a robust understanding of the statistical mechanics low-dimensional inverse problem may shed some light in the general Boltzmann machine learning problem due to the presence of phase transitions for very large systems.

A further reason of interest for the problem we deal with is that, in recent times, this method has attracted some attention due to it's ability to advance a useful novel approach for several applications like neural networks, protein structures, computer vision  \cite{Schneidman2006,Mezard_Mora_2009,Schug22124,Schug22124, MorcosE1293, Geman_Graffigne_1986}, and the socio-economic sciences 
~\cite{McFadden2001,Durlauf1996b,BrockDurlauf2001, Burioni2015EnhancingPT, ContucciVernia2020,BarraContucci2014, GalloBarra2009, BurioniContucci2015, OsabuteyOpokuG2020, OpokuOsabuteyK2019}. 

The system we consider here is made of Ising spins and, beside an homogeneous magnetic field and a constant two body interaction, it contains a constant three body term. One of the peculiarities of this model, which turns out to have a cubic Hamiltonian function, is that it lacks the standard convexity property of its quadratic version and its direct and inverse problems are therefore outside the general methods of convex optimization problems.    
Taking into account the three-body term, we move from a generic graph (network) structure where we consider only dyadic or pairwise interactions into hypergraphs where faces are also considered \cite{BattistonCILLPYP_2020, BattistonAmico2021,MajhiPercGhosh2022}. This allows for the consideration of a large spectrum of applications that are closely related to real-world phenomena, such as team collaborations rather than collaborations between pairs (see \cite{BensonAbebe2018}).
According to \cite{BattistonAmico2021,BensonAbebe2018} the presence of higher-order interactions, such as three or more body interactions, may have significant impact also on the dynamics of interacting networked systems and potentially lead to abrupt transitions between states. Abrupt transitions are a prevalent phenomenon in nature that can be found in everything from social networks to biology \cite{BensonAbebe2018, AlvarezBattiston2021}. 

The model we consider is invariant under the permutation group but its extension to the case in which that symmetry is not present has been already considered in \cite{ContucciKerteszOsabutey2022} with the same perspectives of the multi-populated quadratic models
\cite{FedeleVernia2013, Burioni2015EnhancingPT}. An intriguing feature of such model is that it 
shows a \textit{discontinuous first-order phase transition} which is not present in the case of the standard quadratic mean-field model. 

To solve the inverse problem we first compute, exploiting the exact solution of the model \cite{GO2023,ContucciKerteszOsabutey2022}, the analytical formulas for the system's macroscopic variables in the thermodynamic limit where they provide explicit expressions for the interaction couplings (cubic and quadratic) and the magnetic field. It is worth noticing that since the number of necessary relations to compute the free parameters is three we need to make observations up to the third moment of the probability distribution. To relate the analytical inversion with the (statistical) observations we use the maximum likelihood criteria and we advance a link between estimated and theoretical values. Finally, we test how well the model's free parameters are reconstructed using the inversion formulas and how their robustness is affected by both the system size and the number of independent samples simulated from the model's equilibrium configuration.

The paper is organised as follows. The cubic mean-field model is
introduced in Section \ref{sec2} where it has been shown how to compute and test the robustness of 
the analytical inverse formulas using the maximum likelihood estimation procedure.
Section \ref{sec3} is devoted to the numerical testing of the robustness of the inversion formulas for unique stable solutions. In Section \ref{sec4} the case of metastable or multiple solutions for finite-size systems is discussed. The final section, Section \ref{sec5}, provides a general conclusion and the model's future prospects. 

\section{\label{sec2}Inverse problem for the cubic mean-field Ising model}
\begin{center}
\begin{figure}[h]
\includegraphics[width=8cm]{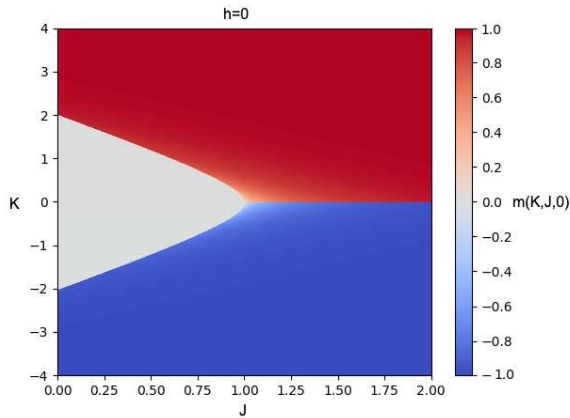}
\caption{\label{PhaseD} $h=0$. Phase diagram of the stable solutions of \eqref{SCc} showing the coexistence curves. For $J<1$, three distinct phases are observed: the negatively polarised phase (in blue), the zero or unpolarized phase (in gray), and the positively polarised phase (in red). As a result, in that region, a progressive increase in K from negative to positive values encounters two consecutive jumps.}
\end{figure}
\end{center}

\begin{figure}%
\centering
\subfigure[Magnetization]{%
\label{fig:magnetization}%
\includegraphics[height=2in]{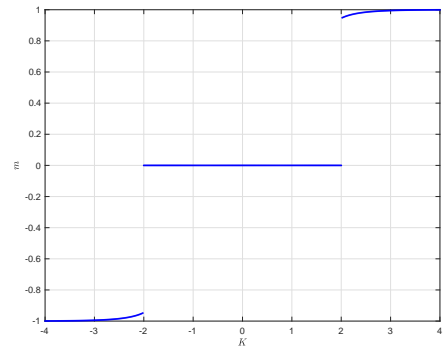}} \
\subfigure[Susceptibility]{%
\label{fig:susceptibility}%
\includegraphics[height=2in]{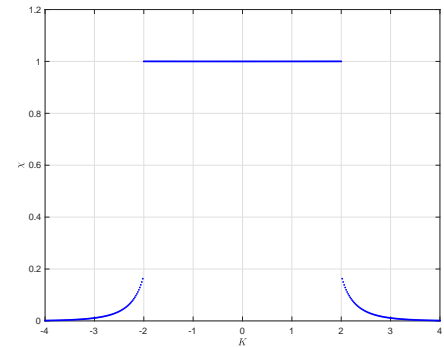}} \
\subfigure[Third moment]{%
\label{fig:3rdmoment}%
\includegraphics[height=2in]{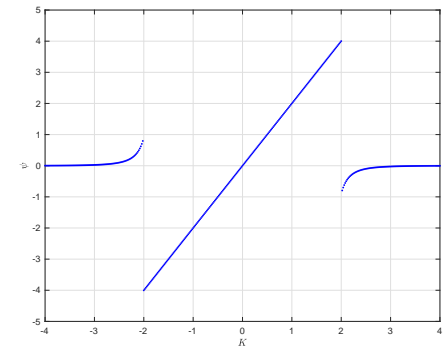}}%
\caption{$J=0$, $h=0$. First three moments of the model as a function of $K$: In \subref{fig:magnetization} the total magnetisation shows indication
of phase transitions occurring at a critical point around $\pm 2$. At the critical point the  susceptibility as seen in \subref{fig:susceptibility} and the third moment in \subref{fig:3rdmoment} has a jump to $1$ and a jump to around $\pm 4$ respectively. 
}
\label{MST_K}%
\end{figure}

Let us consider the Hamiltonian of an Ising model on $N$  spin configurations, $\Omega_N =\{ -1, +1\}^N$, with cubic interaction and spin moments $\sigma_i=\pm1$, $i=1,\ldots, N$, defined as
\begin{equation}\label{3CW}
    H_N(\sigma) = - \sum_{i,j,k = 1}^N K_{i,j,k} \sigma_i\sigma_j\sigma_k - \sum_{i,j = 1}^N J_{i,j} \sigma_i\sigma_j - \sum_{i = 1}^N h_{i} \sigma_i.
\end{equation}
Assuming  mean-field interaction, we set $ K_{i,j,k} = \frac{K}{3N^2},  \; J_{i,j} = \frac{J}{2N} \; \text{and} \; h_i = h$ where $K, J$ are the cubic and binary spin coupling and $h$ is the external magnetic field. Hence, the Hamiltonian per particle is 
\begin{equation}\label{1CCW}
H_N(\sigma) = - N\left(\frac{K}{3} m_N^3(\sigma)+ \frac{J}{2} m_N^2(\sigma) + h m_N(\sigma)\right),
\end{equation}
 where
\begin{equation}\label{mag}
    m_N(\sigma) = \frac{1}{N} \sum_{i = 1}^{N} \sigma_i
\end{equation}
is the magnetisation per particle of the configuration $\sigma$. The Boltzmann-Gibbs state on a configuration $\sigma$ is given by 
\begin{equation}\label{prob}
    P_{N,K,J,h} (\sigma) = \frac{e^{-  H_N(\sigma)}}{Z_N},
\end{equation}
where $Z_N=\sum_{\sigma\in\Omega_N} e^{-  H_N(\sigma)}$ is the partition function of the system. As a result, we obtain the pressure function per particle associated with the thermodynamic system as: 
\begin{equation}\label{FEnergy}
p_N = \dfrac{1}{N}\,\log Z_N  \; .
\end{equation}
For a given observable $f(\sigma)$ the Boltzmann-Gibbs expectation $\omega_N(f(\sigma))$ is defined as follows:
\begin{equation}\label{exp_m}
    \omega_N(f(\sigma)) = \frac{\sum_{\sigma\in\Omega_N} f(\sigma) e^{-H_N(\sigma)}}{Z_N}.
\end{equation}
Furthermore, the pressure function \eqref{FEnergy} can be used to generate the moments of the system with respect to the Boltzmann-Gibbs measure. Hence, one obtains the following finite-size quantities:
\begin{equation}\label{f-omega}
\frac{\partial p_N}{ \partial h} = \omega_N(m_N(\sigma))
\end{equation}
\begin{equation}\label{f-sp}
\frac{\partial ^2 p_N}{ \partial h^2}
=\chi_N=N[\omega_N(m_N^2(\sigma)) - \omega_N^2(m_N(\sigma))] 
\end{equation}
and 

\begin{equation}\label{f-sp2}
\begin{split}
\frac{\partial ^3 p_N}{\partial h^3} = \psi_N 
= N^2[ \omega_N(m^3_N) &- 3\omega_N(m_N)\omega_N(m^2_N)\cr
&+ 2\omega_N^3(m_N)]
\end{split}
\end{equation}
where $\omega_N(m_N(\sigma))$, $\chi_N$ and $\psi_N$  are the finite-size average magnetisation, susceptibility and third moment respectively. The considered model can be solved exactly \cite{GO2023} using the large deviations technique, which was proposed in \cite{Ellis85}. The thermodynamic limit of \eqref{FEnergy} admits the following variational representation \cite{GO2023}:

\begin{equation}\label{var1CCW}
p(K,J,h) = \lim_{N\to\infty} p_N = \sup_{m\in[-1,1]} \phi(m),
\end{equation}
where $\phi(m)= U(m)- I(m)$ with

\begin{equation}\label{energy}
U(m)= \dfrac{K}{3} m^3 +\dfrac{J}{2} m^2 +h m 
\end{equation}
is the energy contribution and

\begin{equation}\label{entropy}
I(m)=\frac{1-m}{2}\log\left(\frac{1-m}{2}\right)+\frac{1+m}{2}\log\left(\frac{1+m}{2}\right)
\end{equation}
is the entropy contribution. The stationarity condition, that acts as a consistency equation, gives

\begin{equation}\label{SCc}
    m=\tanh( Km^2 +  Jm +   h),
\end{equation}
and must be satisfied by the solutions of the variational principle \eqref{var1CCW}. 
In order to solve the inverse problem analytically for a given configuration of spin particles, we first find the relation between the model parameters and the variational principle \eqref{var1CCW}. Observe that,

\begin{equation}\label{consistency eqn}
    \frac{\partial p}{\partial h}=m, \qquad \text{i.e.,} \quad m=\tanh{(Km^2 + Jm + h)},
\end{equation}

\begin{equation}\label{sp}
\frac{\partial^2 p}{\partial h^2} = \chi = \frac{(1-m^2)}{1-(1-m^2)(J+2Km)} \qquad \text{and}
\end{equation}

\begin{equation}\label{sp2}
\frac{\partial^3 p}{\partial h^3} = \psi = \frac{2\chi^2\bigg((K(1 - 3m^2) - Jm)\chi-m\bigg)}{(1-m^2)}.
\end{equation}
The peculiar feature of the cubic mean-field model is the presence of three distinct stable phases in the magnetic order parameter $m$. Unlike the usual quadratic model, here an unpolarised stable phase close to $m=0$ appears beyond the usual two phases of positive and negative magnetization. From Figure \ref{PhaseD} one can observe a triple point $(K,J,h)=(0,1,0)$ where all the three phases meet \cite{ContucciKerteszOsabutey2022}. 
Let us consider the model in its simplest form with zero quadratic coupling and magnetic field i.e. when $J=h=0$ and only the cubic coupling in \eqref{1CCW} is present. It is worth mentioning that when $J=h=0$ and $K$ is progressively increased from negative to positive, one encounters two transitions: from a negatively polarized phase to an unpolarized one and from an unpolarized phase to a positively polarized one (see Fig. \ref{PhaseD}; and Fig $1$. of \cite{ContucciKerteszOsabutey2022}). In Figure \ref{MST_K} we illustrate an example of critical behaviour for our model with the presence of phase transitions occurring at $J=h=0$ when $K$ is varied.

The quantities, $m, \chi$ and $\psi$ are the infinite volume limit average magnetisation, susceptibility and third moment corresponding to the finite-size quantities $\omega_N, \chi_N$ and $\psi_N$ respectively in the thermodynamic limit. The system of equations \eqref{consistency eqn}, \eqref{sp} and \eqref{sp2} has three unknowns $K,J$ and $h$ which can be solved. Having knowledge of $m, \chi$ and $\psi$ one can compute the parameters (i.e. $K,J$ and $h$) of the model through the following equations:

\begin{equation}\label{K_inv}
K = \frac{m}{(1-m^2)^2} + \frac{\psi}{2\chi^3},
\end{equation}  

\begin{equation}\label{J_inv}
    J = \frac{1}{1-m^2} -  \frac{1}{\chi} - 2Km
\end{equation}
and the external magnetic field is then obtained from \eqref{SCc} as
\begin{equation}\label{h_inv}
    h = \tanh^{-1} (m) - Km^2 - Jm.
\end{equation}
Let us observe that, in the region of the parameter space where the consistency equation \eqref{SCc} 
has a unique  stable solution the following holds:
\begin{equation}\label{Exp(mN)}
\lim_{N\rightarrow\infty}  \omega_N(m_N(\sigma)) = m.
\end{equation}
In analogy to the behaviour of the  quadratic case \cite{FedeleVernia2017}, the Boltzmann-Gibbs measure \eqref{prob} may be multimodal for some $(K,J,h)$ in the parameter space for both the finite-size system and in the thermodynamic limit. In this case equation \eqref{Exp(mN)} fails to hold. We will discuss later how to handle such a case, following the work done in \cite{FedeleVernia2017,ContucciLuziV2016}. The procedure discussed so far deals with the analytical inverse problem. The remainder of this section will be devoted to the statistical procedure required to compute the estimators of $K,J$ and $h$.

We start by generating $M$ independent configurations $\sigma^{(1)}, ..., \sigma^{(M)}$ distributed according to \eqref{prob} from the model's equilibrium configuration.
Notice that the analytical inverse formulas of $K,J$ and $h$ in equations \eqref{K_inv}, \eqref{J_inv} and \eqref{h_inv} respectively, are valid on the infinite volume limit of the observables, i.e. $m$, $\chi$ and $\psi$. Hence, to compute the estimates of the model parameters $K,J$ and $h$, the maximum likelihood estimation  procedure will be adopted having knowledge of real data. This procedure ensures that the estimated model parameters maximize the probability of getting the given sample of spin configurations from the distribution. Furthermore, the analytical inverse procedure requires statistical approximation of the infinite volume limit quantities (i.e. $m, \chi$ and $\psi$) which are substituted by their finite-size forms $\omega_N$, $\chi_N$ and $\psi_N$. The maximum likelihood function for the measure \eqref{prob} is defined as 
\begin{equation*}
\begin{split}
    L(K,J,h) &= P_{N,K,J,h}\{\sigma^{(1)}, ..., \sigma^{(M)}\}\cr
    &=\prod_{l=1}^M P_{N,K,J,h}\{\sigma^{(l)}\}\cr
    &=\prod_{l=1}^M \frac{e^{-H_N(\sigma^{(l)})}}{\sum_{\sigma\in\Omega_N} e^{-H_N(\sigma)}}.
\end{split}
\end{equation*}
This procedure will enable defining the finite-size magnetisation $\omega_N(m_N(\sigma))$ in terms of the empirical average (i.e. $m_N$) for each of the $M$ sampled spin configurations. Further, we have that
\begin{equation}
    \ln L(K,J,h) = \sum_{l=1}^M\left[(-H_N(\sigma^{(l)})) - \ln \sum_{\sigma\in\Omega_N}e^{-H_N(\sigma)} \right].
\end{equation}
The derivatives with respect to the parameters $K,J$ and $h$ are given below as:
\begin{equation*}
\begin{split}
  \frac{\partial}{ \partial h} \ln L(K,J,h) &= N\sum_{l=1}^M\left(m_N(\sigma^{(l)}) - \omega(m_N(\sigma))\right)\cr
  \frac{\partial}{ \partial J} \ln L(K,J,h) &= \frac{N}{2}\sum_{l=1}^M\left(m_N^2(\sigma^{(l)}) - \omega(m_N^2(\sigma))\right)\cr
    \frac{\partial}{ \partial K} \ln L(K,J,h) &= \frac{N}{3}\sum_{l=1}^M\left(m_N^3(\sigma^{(l)}) - \omega(m_N^3(\sigma))\right)
\end{split}
\end{equation*}
and they vanish when

\begin{equation}\label{moments}
\begin{split}
  \omega_N(m_N(\sigma)) &= \frac 1M\sum_{l=1}^M m_N(\sigma^{(l)})  \cr
  \omega_N(m_N^2(\sigma)) &= \frac 1M\sum_{l=1}^M m_N^2(\sigma^{(l)}) \cr
  \omega_N(m_N^3(\sigma)) &= \frac 1M\sum_{l=1}^M m_N^3(\sigma^{(l)}).
\end{split}
\end{equation}
The function $L(K,J,h)$ is at its maximum when the 
first, second and third moments of the magnetization in equation \eqref{moments} are obtained. It is worth noticing that 

\begin{equation}
    m_N(\sigma^{(l)}) = \frac{1}{N} \sum_{i=1}^N \sigma_i^{(l)} \qquad \text{for} \quad l = 1,\ldots,M
\end{equation}
are the total magnetizations of the $M$ sample spin configurations.
Now, the inverse problem can be solved when we make use of \eqref{K_inv}, \eqref{J_inv}, \eqref{h_inv} and \eqref{moments}. 
The maximum likelihood procedure computes the estimators of the infinite volume quantities $m$, $\chi$ and $\psi$, from a sample data set through the following:

\begin{equation}\label{ML_omega}
    \widehat{m} = \frac{1}{M} \sum_{l=1}^M m_N(\sigma^{(l)}),
\end{equation}
\begin{equation}\label{ML_chi}
   \widehat{\chi} = N\bigg(\frac{1}{M} \sum_{l=1}^M m_N^2(\sigma^{(l)})-\widehat{m}^2\bigg)
\end{equation}
and 
\begin{equation}\label{ML_psi}
 \widehat{\psi} = N^2\left(\frac{1}{M} \sum_{l=1}^M m^3_N(\sigma^{(l)}) - 3\widehat{m}\frac{1}{M} \sum_{l=1}^M m^2_N(\sigma^{(l)}) + 2\widehat{m}^3\right).
\end{equation}

We now define the estimators of the three parameters of the cubic mean-field model using the statistical estimators for the magnetization, susceptibility and third moment \eqref{ML_omega}, \eqref{ML_chi}  and \eqref{ML_psi} in the infinite volume limit relations among those quantities \eqref{K_inv}, \eqref{J_inv}
 and \eqref{h_inv}

\begin{equation}\label{Kexp}
    \widehat{K} = \frac{\widehat{m}}{(1-\widehat{m}^2)^2} + \frac{\widehat{\psi}}{2\widehat{\chi}^3}, 
\end{equation}

\begin{equation}\label{Jexp}
    \widehat{J} = \frac{1}{1-\widehat{m}^2} - \frac{1}{\widehat{\chi}} - 2\widehat{K} \widehat{m},
\end{equation}

and 

\begin{equation}\label{hexp}
    \widehat{h} = \tanh^{-1}(\widehat{m}) - \widehat{K}\widehat{m}^2 - \widehat{J} \widehat{m} .
\end{equation}

At the critical point $(K,J,h)=(0,1,0)$ where all the three phases meet the magnetization is zero and the infinite volume magnetic susceptibility $\chi$ and the third moment $\psi$ defined by equations \eqref{sp} and \eqref{sp2} respectively diverge.  Hence, the inversion formulas \eqref{K_inv}, \eqref{J_inv} and \eqref{h_inv} does not hold as it will be illustrated at the end of the next section. We do not include the inversion formulas at the critical point in this work but the problem will be considered in future work.

\section{\label{sec3}Test for the case of unique solution}
\begin{center}
\begin{figure}[h]
\includegraphics[width=24em]{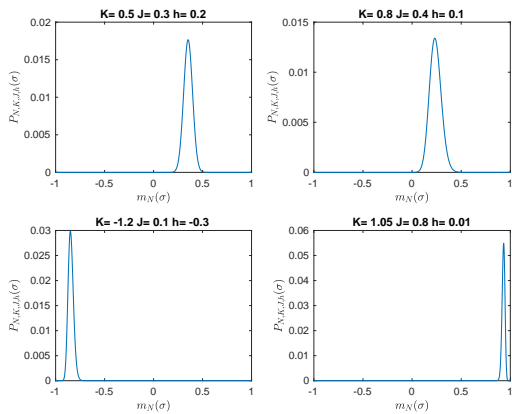}
\caption{\label{StaBoltz} Boltzmann Gibbs distribution of the total magnetisation for $N=1000$ and different set of triples ($K,J,h$). }
\end{figure}
\end{center}

In this section we are going to examine how the inversion equations perform for different and increasing choices of $N$ and $M$, respectively the number of particles and sampled configurations. 
The specific case we consider is the inversion problem for those values of the triple ($K,J,h$) where  there is a unique stable solution of \eqref{SCc}. In this case, the Boltzmann-Gibbs distribution of the total magnetisation has a unique peak always centered around the analytic solution $m$: some examples are shown in Figure \ref{StaBoltz} for fixed $N$. The accuracy of the estimation increases as $N$ and $M$ increase. 

\begin{center}
\begin{figure}[h]
\includegraphics[width=28em]{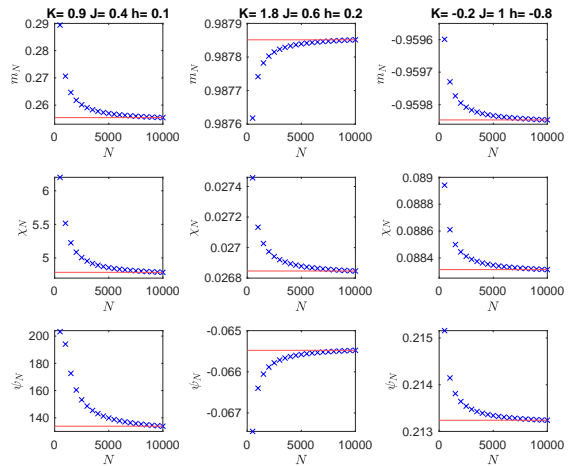}
\caption{\label{StabMST} Finite-size average magnetization $m_N$, susceptibility $\chi_N$ and third moment $\psi_N$ as functions of $N$ for three different set of triples ($K,J,h$). Blue crosses represent the values of $m_N$ (upper panels), $\chi_N$ (middle panels) and $\psi_N$ (lower panels) for varying $N$. As $N$ increases $m_N$,  $\chi_N$ and $\psi_N$ approach their true values in the thermodynamic limit given as the red horizontal lines for the chosen values of $K,J$ and $h$.}
\end{figure}
\end{center}

The parameters $K,J$ and $h$ are obtained from the computation of the finite-size quantities $m_N, \chi_N$ and $\psi_N$  using configurations extracted from the Boltzmann-Gibbs distribution of the data. Estimation of $m_N$, $\chi_N$ and $\psi_N$ for fixed triples of the parameters ($K,J,h$) and varying $N\in[500,10000]$ are shown in Figure \ref{StabMST}. In the same figure, the thermodynamic limits of those quantities are also shown.

\begin{center}
\begin{figure}[h]
\includegraphics[width=9cm]{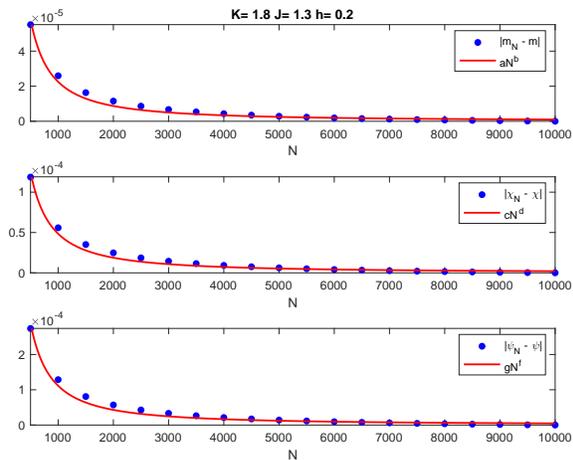}
\caption{\label{Errors} $K=1.8, J=1.3,h=0.2$. Absolute error of the finite size quantities $m_N, \chi_N$ and $\Psi_N$ as functions of $N$ together with the best power law fits. In the upper panel, $|m_N-m|$ is shown as a function of $N$ together with the best fit $a N^b$, where $ a = 0.28 \in (0.06, 0.50)$ and $b = -1.37 \in (-1.49, -1.25)$ with a goodness of fit $R^2=0.9829$. The middle panel displays $|\chi_N-\chi|$ as a function of $N$ together with its corresponding best fit $c N^d$, with $c = 0.62 \in (0.14,1.09)$, $ d= -1.37 \in (-1.49,-1.25)$ and $R^2= 0.9830$ as goodness of fit. The lower panel represents $|\psi_N-\psi|$ as a function of $N$ together with its corresponding best fit $g N^f$, with $g = 1.47 \in (0.32,2.62)$, $f = -1.37 \in (-1.49,-1.25)$ and a goodness of fit $R^2 =0.9826$.}
\end{figure}
\end{center}

From Figure \ref{StabMST} we can observe the monotonic behaviour of $m_N$, $\chi_N$  and $\psi_N$ as $N$ increases. In Figure \ref{Errors} we study the relationship between the absolute difference of the finite-size quantities and their corresponding thermodynamic values as a function of the system size $N$. We find evidence that the finite-size quantities $m_N, \chi_N$ and $\psi_N$ converge to their true values with a power law behaviour as $N$ increases. The obtained results indicate that using $N=10 000$ one can estimate the infinite volume magnetisation, susceptibility and the third moment with vanishing error. We will proceed to use $N=10 000$ as the size for each of the $M$ independent spin configurations $\sigma^{(1)}, ..., \sigma^{(M)}$. Further numerical tests will be performed to determine a suitable number of sample configurations $M$ that can be used for reconstructing the model parameters using the inversion formulas. 

\begin{center}
\begin{figure}[h]
\includegraphics[width=28em]{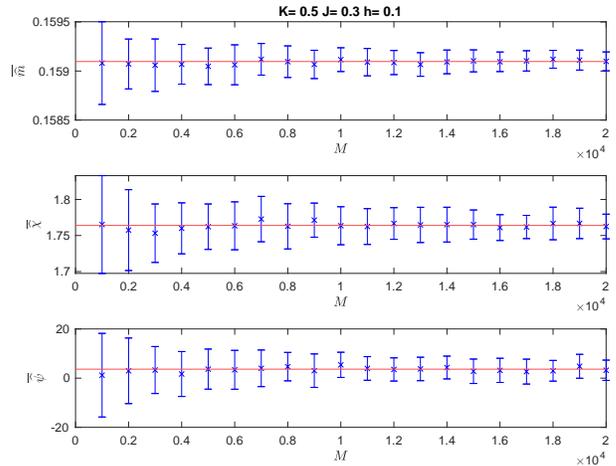}
\caption{\label{StabNS_MST} $K=0.5, J=0.3, h=0.1.$ Reconstructed average magnetization $\overline{\widehat{m}}$, susceptibility $\overline{\widehat{\chi}}$ and third moment $\overline{\widehat{\psi}}$ (blue crosses) as a function of $M$ with standard deviation on $50$ different $M$-sample and $N = 10 000$. The continuous red line corresponds to $m$,  $\chi$ and $\psi$ in the thermodynamic limit.}
\end{figure}
\end{center}

\begin{center}
\begin{figure}[h]
\includegraphics[width=28em]{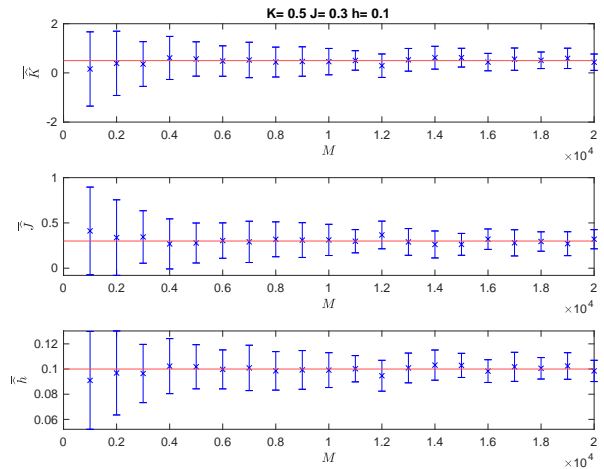}
\caption{\label{StabNS_KJh} $K=0.5, J=0.3, h=0.1.$ $\overline{\widehat{K}}, \overline{\widehat{J}}$ and  $\overline{\widehat{h}}$ as a function of $M$ for $N = 10000$. The blue crosses are the estimation of $\overline{\widehat{K}}, \overline{\widehat{J}}$ and $\overline{\widehat{h}}$ with standard deviations on $50$ different $M$-samples of configurations of the same system. The horizontal red line in each panel corresponds to the exact values of $K, J$ and $h$.}
\end{figure}
\end{center}

To obtain the standard deviations associated to the reconstruction of the estimators, we simulate from the model's equilibrium configuration $50$ different instances of the $M-\mathrm{iid}$ sample configurations, i.e. ($\sigma^{(1)},\ldots,\sigma^{(M)}$), apply the maximum likelihood estimation procedure to each of them separately, solve the inverse problem using \eqref{Kexp}, \eqref{Jexp} and \eqref{hexp} and then  average the inferred values over the $50$ different $M$-samples. The mean value of the estimators $\widehat{m}, \widehat{\chi}, \widehat{\psi}$,  and  ($\widehat{K},\widehat{J},\widehat{h}$) over the $50$ different $M$-samples of spin configurations are denoted by  $\overline{\widehat{m}}, \overline{\widehat{\chi}}, \overline{\widehat{\psi}}$,  and  ($\overline{\widehat{K}},\overline{\widehat{J}},\overline{\widehat{h}}$) respectively. The results are shown in Figures \ref{StabNS_MST} and \ref{StabNS_KJh}. 
Figures \ref{StabNS_MST} and \ref{StabNS_KJh} illustrate that at $M=20 000$ we get smaller error bounds for the reconstruction as compared to lesser values of $M$. 

\begin{center}
\begin{figure}[h]
\includegraphics[width=25em]{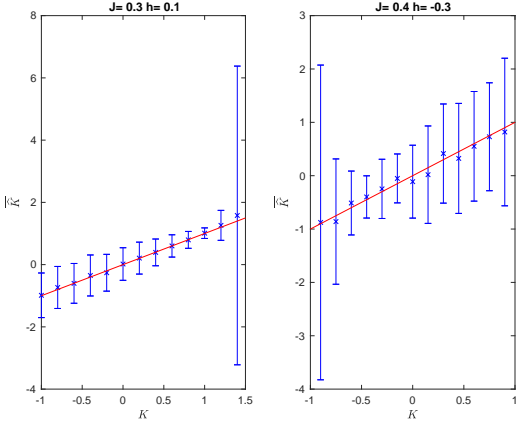}
\caption{\label{StabV_K}  $\overline{\widehat{K}}$ as a function of $K$ for $N=10000$ and $M = 20 000$. $J=0.3, h=0.1$ in left panel and $J=0.4, h=-0.3$ in the right panel. The estimations of $\overline{\widehat{K}}$ are given as the blue crosses in both panels with standard deviations on 50 different $M$-samples of configurations of the same system. The red continuous line represents $\overline{\widehat{K}} = K$. }
\end{figure}
\end{center}

\begin{center}
\begin{figure}[h]
\includegraphics[width=26em]{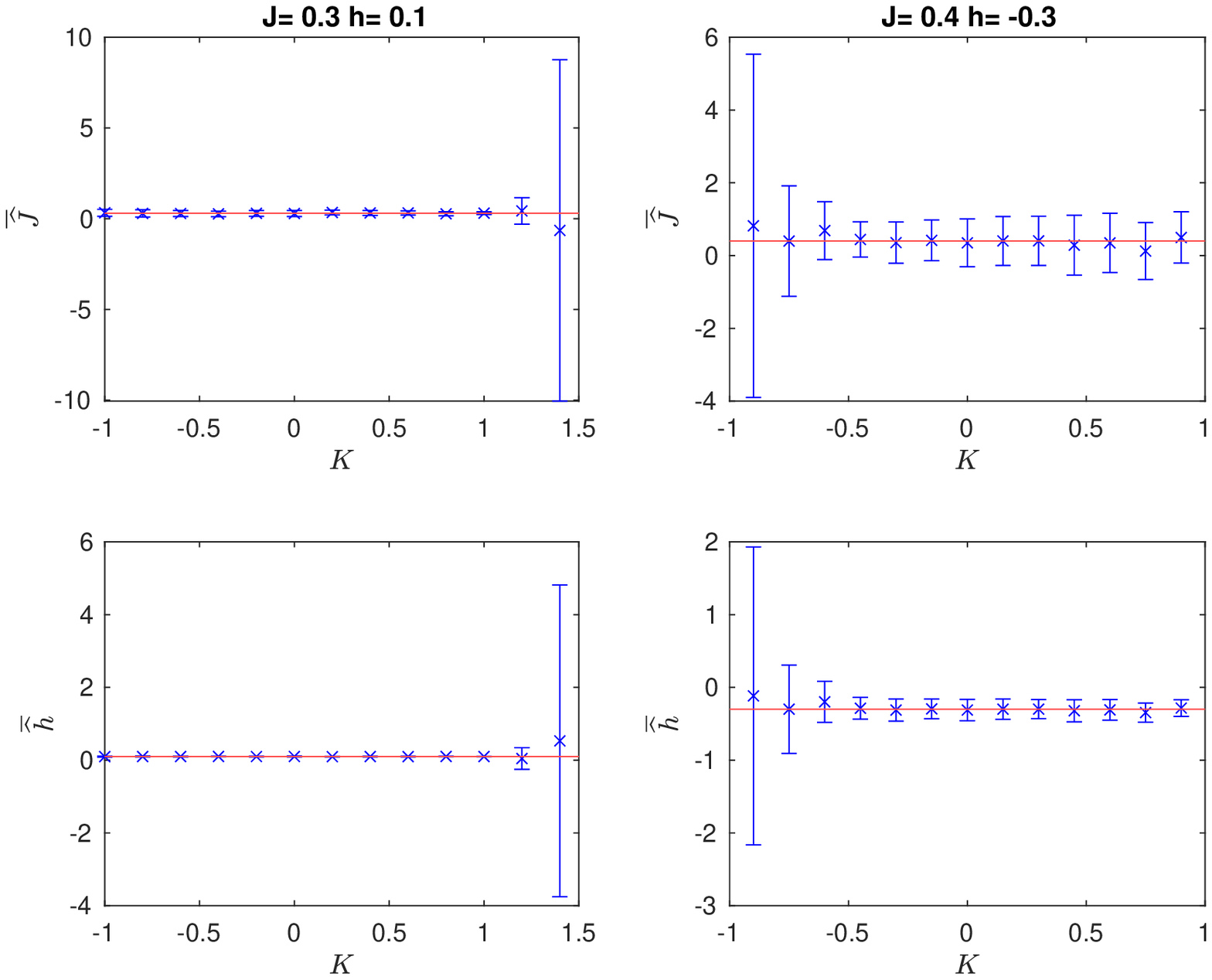}
\caption{\label{StabV_KJh} $\overline{\widehat{J}}$ and $\overline{\widehat{h}}$ as a function of $K$ for $N=10000$ and $M = 20 000$. $J=0.3, h=0.1$ in the left panels and $J=0.4, h=-0.3$ in the right panels. The estimates of $\overline{\widehat{J}}$ and $\overline{\widehat{h}}$ are given as the blue crosses in all the panels with standard deviations on 50 different $M$-samples of configurations of the same system. The red continuous lines represent the exact values of $J$ and $h$.}
\end{figure}
\end{center}

\begin{center}
\begin{figure}[h]
\includegraphics[width=24em]{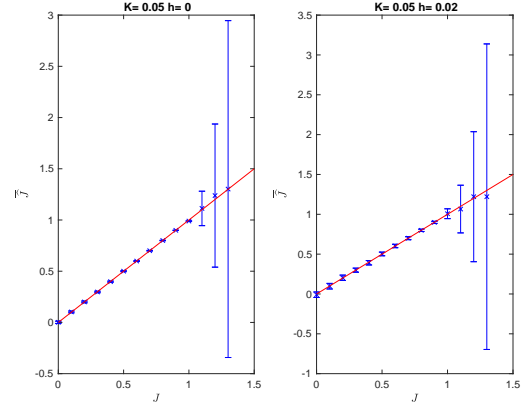}
\caption{\label{StabV_J1} $\overline{\widehat{J}}$ as a function of $J$ for $N=10000$ and $M = 20 000$. $K=0.05, h=0$ in the left panel and $K=0.05, h=-0.02$ in the right panel. The blue crosses are the reconstructed values of $J$ in both panels with standard deviations on 50 different $M$-samples of configurations of the same system. The red continuous line  represents the exact value $\overline{\widehat{J}}=J$. }
\end{figure}
\end{center}

\begin{center}
\begin{figure}[h]
\includegraphics[width=26em]{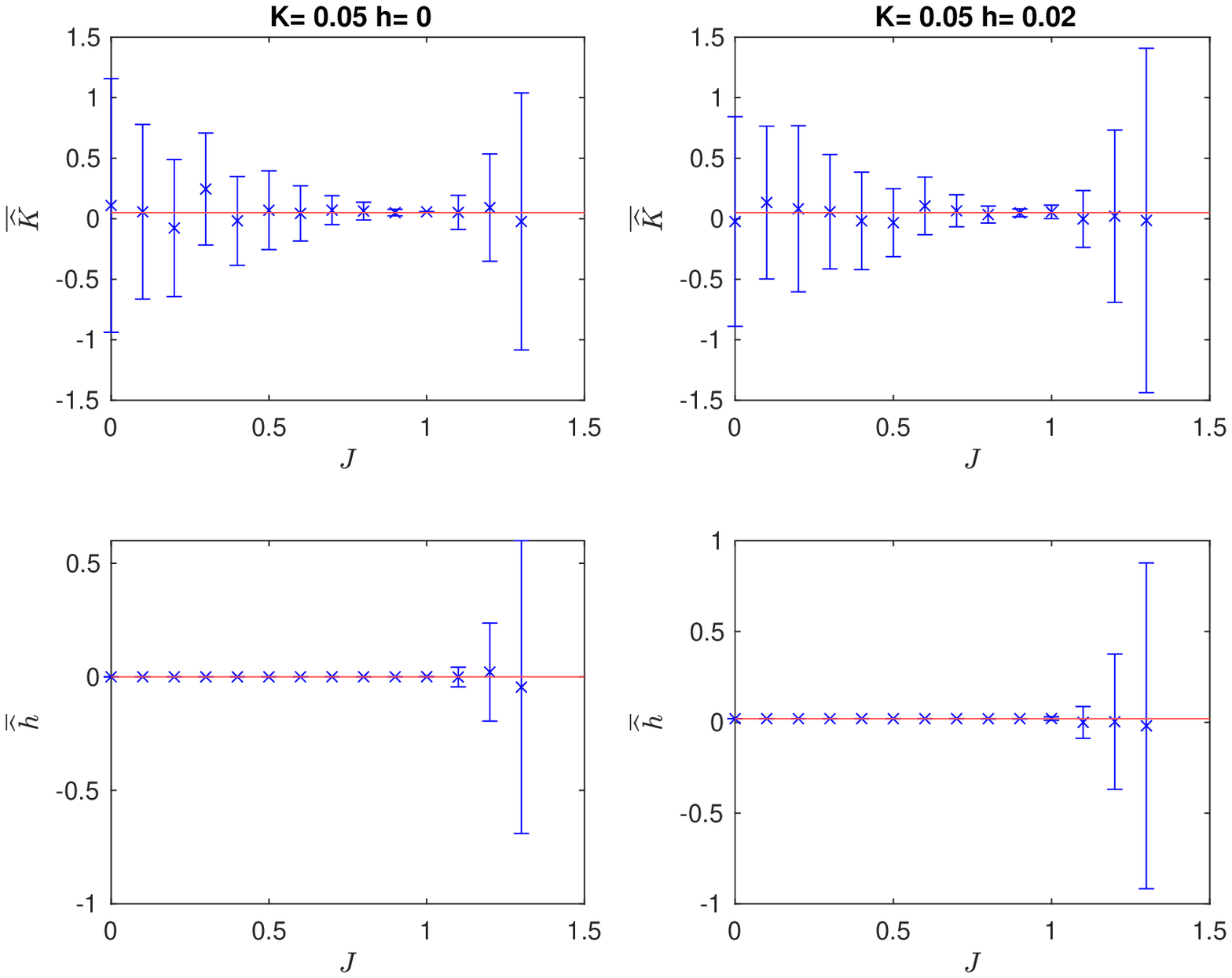}
\caption{\label{StabV_JKh1} $\overline{\widehat{K}}$ and $\overline{\widehat{h}}$ as a function of $J$ for $N=10000$ and $M = 20 000$. $K=0.05, h=0$ in the left panels and $K=0.05, h=0.02$ in the right panels. The estimate of $\overline{\widehat{K}}$ and $\overline{\widehat{h}}$ are given as the blue crosses in all the panels with standard deviations on 50 different $M$-samples of configurations of the same system. The red continuous lines represent the exact values of $K$ and $h$.}
\end{figure}
\end{center}

In the sequel, we study the behaviour of the reconstructed parameter for fixed values of $J$ and $h$ and varying $K$ (Figures \ref{StabV_K} and \ref{StabV_KJh}) and also for fixed values of $K$ and $h$ and varying $J$ (Figures \ref{StabV_J1} and \ref{StabV_JKh1}). The simulations are performed using $M = 20 000$, $N=10 000$ and error bars are standard deviations on $50$ different $M$-samples of the same system. We find all the reconstructed parameter values in good agreement with the exact ones. We can observe that as the intensity of the cubic and quadratic coupling increases the error bars associated to the reconstructed parameters grow, as we can expect since in that region of the parameter space the system is more disordered due to the presence of multiple local stable states and the fluctuations are greater.

\begin{center}
\begin{figure}[h]
\includegraphics[width=28em]{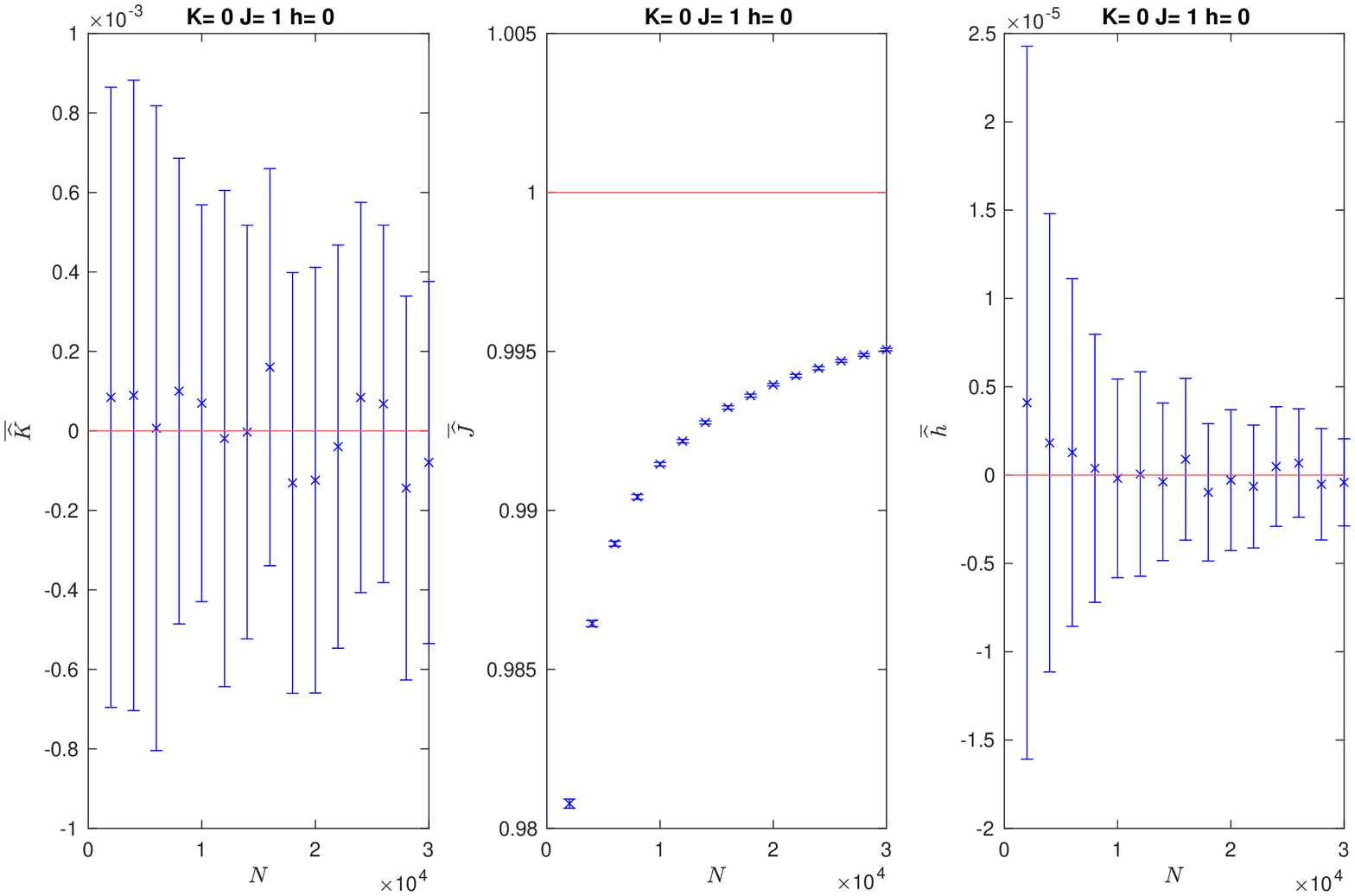}
\caption{\label{Stab_V_NKJh} $K=0, J=1, h=0$. $\overline{\widehat{K}}, \overline{\widehat{J}}$ and $\overline{\widehat{h}}$ as a function of $N$ for $M=20000$.  The reconstructed estimates of $K,J$ and $h$ are given as the blue crosses on statistical error bars of $50$ different $M-$samples. The red continuous line is the exact value of the parameters $K,J$ and $h$ in the respective panels.}
\end{figure}
\end{center}

Furthermore, Figure \ref{Stab_V_NKJh} show the reconstructed parameters as a function of $N$ at the critical point ($K=0, J=1, h=0$). It can be noticed that the reconstruction at the critical point for $K$ and $h$ agrees with their exact values with only a small percentage of error and that of $J$ is underestimated. 


It worth observing that when $K=h=0$ and $J>1$ the consistency equation \eqref{SCc} has two stable solutions. In this case, for the finite-size system and in the thermodynamic limit, the Boltzmann-Gibbs distribution of the total magnetization presents two peaks each centered around one of the stable solutions. In such a case the inverse problem procedure discussed in Section \ref{sec2} cannot be used for the reconstruction of the model parameters. We refer readers to \cite{FedeleVernia2013} where this case has been studied using the spin flip approach due to symmetry of the solution in both finite-size and infinite volume systems for the quadratic mean-field model. The clustering algorithm to be outlined in the next section provides a more general approach to handle the reconstruction of the model parameters when the phase space has multiple locally stable solution.
%
\section{\label{sec4} Clustering algorithm for metastable state solutions}
Here, we focus on cases where equation \eqref{SCc} has a metastable solution. 
This corresponds to the case where there are more than one locally stable solution of  the consistency equation \eqref{SCc}. For this model, equation \eqref{SCc} can have at most three solutions and $\phi$ has at most two local maxima for 
fixed $(K,J,h)$. The existence of the metastable solution in the infinite volume limit is represented at finite $N$ by the occurrence of an extra peak in the distribution. Therefore, while in the thermodynamic limit the Boltzmann-Gibbs distribution of the magnetisation  
is unimodal with the peak corresponding to the stable solution, in the finite size case also the peak corresponding 
to the metastable one is present and the distribution is bimodal. Hence, in this case, the inversion problem 
cannot be studied globally, as done in the previous section. Instead, the procedure has to be applied locally,
that is to each subset of configurations clustered around the two local maxima.
Given $M$ spin configurations, $\sigma^{(1)}, ..., \sigma^{(M)}$, we perform the reconstruction 
by first partitioning the $M$ configurations in clusters according to their local densities around each local maximum. 
More precisely, using the clustering algorithm discussed in 
\cite{Itoy1993,DecelleR2016,RodriguezL2014,ContucciLuziV2016,ChauBerg2012} 
we divide the $M$ configurations into different clusters using the mutual distances between 
their magnetizations of each configuration. Configurations form a cluster if the magnetization distances are less than a fixed threshold $d_c$. The choice of the optimal threshold 
is obviously crucial: a too small threshold will produce
too many clusters, while a too large one will give only one cluster. Given 
$d_c$, for each configuration $l$ the algorithm computes two quantities: the local density $\rho_l$, defined 
as the number of magnetizations within the given distance $d_c$ to the magnetization of $\sigma^{(l)}$, and the 
minimum distance $\delta_l$ between the magnetization of configuration $l$ and any other configuration with 
a higher density.

The algorithm is based on the assumptions that the cluster centers are surrounded by points with a lower
density, and that the centers are at a relatively large distance from each other. For each configuration, plotting 
the minimum distance $\delta$ as a function of the local density $\rho$ provides a decision graph that gives 
the cluster centers: 
the cluster centers are the outliers in the graph. Finally, each remaining configuration is assigned to the same 
cluster of its nearest neighbor of higher density.
In this study, we identify two clusters $C_k$, $k=1,2$, using the optimal threshold $d_c=0.001$. Notice that it is not 
possible to observe three clusters in the inverse problem due to the analytical properties of the consistency 
equation \eqref{SCc}.

Then, for each cluster $C_k, k=1,2$ we compute the estimates of the finite-size quantities, $\widehat{m}$, $\widehat{\chi}$ and $\widehat{\psi}$, and the corresponding $\widehat{K}, \widehat{J}, \widehat{h}$. More precisely, we can define the estimators of the finite-size quantities  with reference to the clusters as follows:
\begin{equation}
 \widehat{m}_{C_k} = \frac{1}{M_k} \sum_{l\in C_k} m_N(\sigma^{(l)}),
\end{equation}
\begin{equation}\label{ML_chi_e}
  \widehat{\chi}_{C_k} = N\Bigg(\frac{1}{M_k} \sum_{l\in C_k} m_N^2(\sigma^{(l)})-{\widehat{m}_{C_k}}^2\Bigg)
\end{equation}
and 
\begin{widetext}
\begin{equation}\label{ML_psi_e}
\widehat{\psi}_{C_k} = N^2\left(\frac{1}{M_k} \sum_{l\in C_k} m^3_N(\sigma^{(l)}) - 3\widehat{m}_{C_k}\frac{1}{M_k} \sum_{l\in C_k} m^2_N(\sigma^{(l)}) + 2{\widehat{m}_{C_k}}^3\right),
\end{equation}
\end{widetext}
where $M_k$ is the size of the cluster $C_k$, $k=1,2$ such that $M_1 + M_2 = M$. After obtaining the quantities 
above, we now compute the estimated values, $\widehat{K}_{C_k}, \widehat{J}_{C_k}, \widehat{h}_{C_k}$, 
using equations \eqref{Kexp}, \eqref{Jexp} and \eqref{hexp} for each cluster and compute the final estimates 
of the parameters $K$, $J$ and $h$ as the weighted averages:
\begin{equation}\label{clus_K}
\widehat{K} = \frac{1}{M} \sum_{k=1}^2 M_k \widehat{K}_{C_k},
\end{equation}
\begin{equation}\label{clus_J}
\widehat{J} = \frac{1}{M} \sum_{k=1}^2 M_k \widehat{J}_{C_k}
\end{equation}
and 
\begin{equation}\label{clus_h}
\widehat{h} = \frac{1}{M} \sum_{k=1}^2 M_k \widehat{h}_{C_k}.
\end{equation}
Observe that if a point $(K,J,h)$ in the parameter space corresponds to a metastable solution (at finite volume) and it is sufficiently distant from the coexistence curve, we can expect a better reconstruction of the parameters by applying equations \eqref{Kexp}, \eqref{Jexp} and \eqref{hexp} to the configurations in the largest cluster. However, if the point $(K,J,h)$ is close to the coexistence curve, a better reconstruction can be expected using the density clustering algorithm, i.e. by using \eqref{clus_K}, \eqref{clus_J} and \eqref{clus_h}. 

\begin{center}
\begin{figure}[h]
\includegraphics[width=26em]{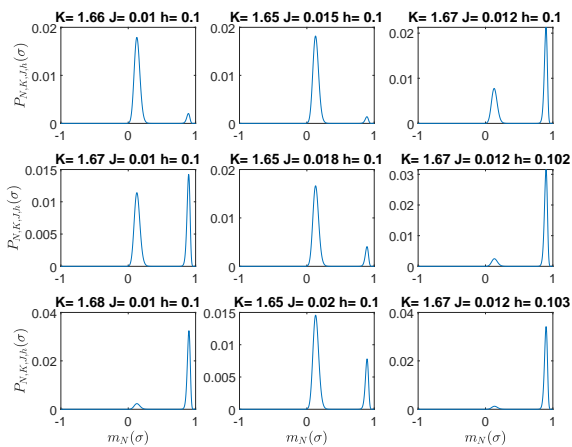}
\caption{\label{MetaStaBoltz} Boltzmann-Gibbs distribution of the total magnetisation with metastable states for fixed $K,J, h$ at $N=1000$. The peaks of the distribution are centered around the two solutions of the consistency equation. }
\end{figure}
\end{center}

Figure \ref{MetaStaBoltz} illustrates how the Boltzmann-Gibbs measure of the magnetization is changing with varying $K,J$ and $h$ in each column starting from the left respectively.

\subsection{Test for metastable state solutions}
\begin{center}
\begin{figure}[h]
\includegraphics[width=24em]{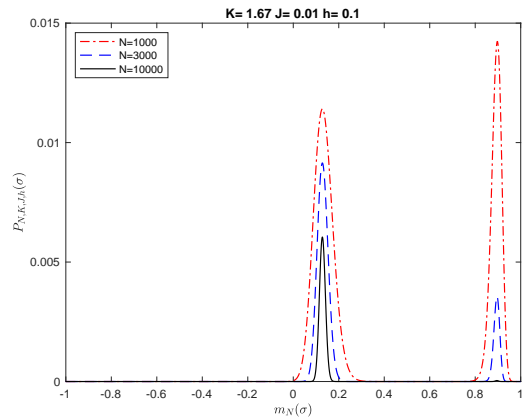}
\caption{\label{Clus_Metastable} $K=1.67, J=0.01, h=0.1$. Boltzmann-Gibbs distribution of the total magnetization at fixed values of $N$. The peaks of the distribution are centered around the two solutions of the equation \eqref{SCc}, with $m_1 = 0.1311$ being the stable solution and $m_2 =0.8973$ the metastable solution. We can observe that the probability of the metastable solution vanishes to $0$ as $N$ goes to infinity (black continuous curve). The red dot-dashed line corresponds to the distribution for $N = 1000$, blue dashed line corresponds to the distribution for $N=3000$ and the black continuous line for the distribution with $N = 10000$.}
\end{figure}
\end{center}

The inverse problem is solved using the density clustering algorithm as discussed and identifying a suitable number of samples $M$ for better reconstruction of the model parameters. The test is performed with $M=20000$ and standard deviations are computed over $50$ different $M$-samples from the same distribution. As an example, consider the  reconstruction of the parameter values $(K,J,h) = (1.67,0.01,0.1)$ for $M=20000$ and $N=3000$. The distribution of the magnetization at this point is given as the blue dashed curve in Figure \ref{Clus_Metastable}, where the two peaks are centered around $m_1 = 0.1311$ and $m_2 =0.8973$, the stable solution and the metastable solution of the consistency equation \eqref{SCc}, respectively.

As is evident from Figure \ref{Clus_Metastable}, the cluster centered around $m_1$ (i.e. $C_1$) has more configurations as compared to the other cluster centered around $m_2$ (i.e. $C_2$). We get the following reconstructed estimates for the parameter values by applying equations \eqref{Kexp}, \eqref{Jexp} and \eqref{hexp} to the setups in both clusters (i.e. $C_1$ and $C_2$) according to formulas  \eqref{clus_K}, \eqref{clus_J} and \eqref{clus_h}:

\begin{equation*}
(\overline{\widehat{K}}, \overline{\widehat{J}}, \overline{\widehat{h}}) = ( 1.76\pm 0.67,-0.11\pm 1.11 ,0.15\pm 0.49).
\end{equation*}
Instead, we obtain the following reconstructed parameter values by applying equations \eqref{Kexp}, \eqref{Jexp} and \eqref{hexp} just to the configurations in the more dense cluster $C_1$: 

\begin{equation*}
(\overline{\widehat{K}}, \overline{\widehat{J}}, \overline{\widehat{h}}) = ( 1.69\pm 0.23,0.01\pm 0.06,0.10\pm 0.004 ).
\end{equation*}
Note that, the reconstructed parameters using only the configurations in the more dense cluster are in better agreement with the exact ones when compared to the reconstructed parameters on both clusters. This is an indication that the point $(K,J,h) = (1.67,0.01,0.1)$ is sufficiently distant from the coexistence curve. Observe that if two clusters have the same density, we do not choose between them and the clustering algorithm provides an optimal reconstruction.

\begin{center}
\begin{figure}[h]
\includegraphics[width=28em]{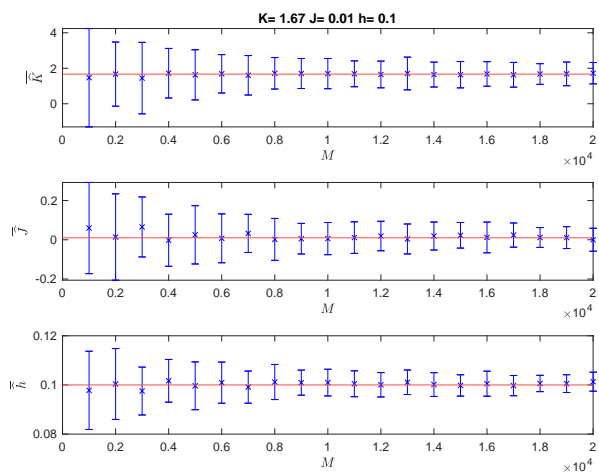}
\caption{\label{ClusNS_KJh} $K = 1.67, J=0.01$ and $h = 0.1$. $\overline{\widehat{K}}, \overline{\widehat{J}}$ and  $\overline{\widehat{h}}$ as a function of $M$ using the largest cluster and $N = 3000$. The reconstructed estimates, $\overline{\widehat{K}}, \overline{\widehat{J}}$ and  $\overline{\widehat{h}}$, are blue crosses on statistical error bars on $50$ different $M$-samples of configurations of the same system. The horizontal red lines in each panel correspond to the exact values of $K,J$ and $h$.}
\end{figure}
\end{center}

Now, we perform reconstruction of the parameters using the cluster with largest size for fixed values of the model parameters and observe its performance for varying $M$ in Figure  \ref{ClusNS_KJh}. It can be observed that the reconstructed parameters are in good agreement to their corresponding exact values. 

As a last remark, note that, given a point $(K,J,h)$ in a neighbourhood of
the coexistence curve, one can observe a metastable state when the number of
particles $N$ is not large enough.  In this case, the clustering algorithm
is useful to  reconstruct the parameters, but it has a high computational
cost.  This is easily overcome by using large number of particles, which
cause the metastable state to vanish (see Figure \ref{Clus_Metastable}) and
the inversion formulas in equations \eqref{Kexp}, \eqref{Jexp}, \eqref{hexp}
become efficient.

\section{\label{sec5}Conclusion}
In this work we consider a mean-field statistical mechanics model with three-body interaction displaying a first order phase transition. We studied and solved the inverse problem and tested the statistical robustness of the inversion method. We numerically tested the inversion method for cases where the consistency equation \eqref{SCc} has a unique stable solution as well as more than one locally stable solution. For the case where the consistency equation \eqref{SCc} has multiple locally stable solution, we used the clustering algorithm to reconstruct the model parameters.

Robustness was tested for different values of the number of particles $N$ and samples $M$ and reached the precision of a few percent for $M=2\times 10^{4}$. 
%
%
We plan to investigate in the future two extensions of the inverse problem: first to the critical point where some of the observables, such as $\chi$ and $\psi$, diverge and to the multi-populated version of the model that found applications  to the description of human-AI ecosystems \cite{ContucciKerteszOsabutey2022}.

\begin{acknowledgments}
The authors thank Claudio Giberti and Emanuele Mingione for useful discussions and G.O. appreciates Filippo Zimmaro for interesting discussions. 
\end{acknowledgments}

\bibliography{apssamp}

\end{document}